# Performance analysis for OFDM-based multi-carrier continuous-variable quantum key distribution with arbitrary modulation protocol

HENG WANG, YAN PAN, YUN SHAO, YAODI PI, TING YE, YANG LI, TAO ZHANG, JINLU LIU, JIE YANG, LI MA, WEI HUANG[†] AND BINGJIE XU*

*Science and Technology on Communication Security Laboratory, Institute of Southwestern Communication, Chengdu 610041, China*
†*huangwei096505@aliyun.com,\*xbjpku@163.com*

**Abstract:** Multi-carrier continuous-variable quantum key distribution (CV-QKD) is considered to be a promising way to boost the secret key rate (SKR) over the existing single-carrier CV-QKD scheme. However, the extra excess noise induced in the imperfect multi-carrier quantum state preparation process of $N$ subcarriers will limit the performance of the system. Here, a systematic modulation noise model is proposed for the multi-carrier CV-QKD based on the orthogonal frequency division multiplexing (OFDM). Subsequently, the performance of multi-carrier CV-QKD with arbitrary modulation protocol (e.g. QPSK, 256QAM and Gaussian modulation protocol) can be quantitatively evaluated by combining the security analysis method of the single-carrier CV-QKD. Under practical system parameters, the simulation results show that the SKR of the multi-carrier CV-QKD can still be significantly improved by increasing the carrier number $N$ even with imperfect practical modulations. Specifically, the total SKR of multi-carrier CV-QKD can be optimized by carefully choosing $N$. The proposed model provides a feasible theoretical framework for the future multi-carrier CV-QKD experimental implementation.

## 1. Introduction

Quantum key distribution (QKD) allows to distribute an information-theoretically secure key between two legal communication parties through an insecure channel based on the fundamental principles of quantum mechanics [1-4]. Compared with the intensively developed discrete-variable QKD (DV-QKD) based on single-photon detection, continuous-variable QKD (CV-QKD) based on coherent detection is compatible with commercial off-the-shelf components and has the potential advantages of high secret key rate (SKR) in metropolitan area [5-9].

Currently, single-frequency optical carrier scheme is regularly used in most existing CV-QKD setups. The SKR of the single-carrier scheme can be improved by increasing the system repetition rate [10-12] and optimizing the excess noise [13-15]. However, the repetition rate of the state-of-art CV-QKD system has reached 5 GHz [10], where the technical challenge lies in that it requires faster data acquisition cards with high linearity, wider bandwidth homodyne detectors with low noise and faster post-processing with high complexity [16-20]. Meanwhile, the excess noise of CV-QKD system has been controlled to a reasonably low level at the order of 0.001~0.01 [10, 15]. To further improve the SKR, the multi-carrier CV-QKD scheme has been proposed to distribute multiplexing independent secret keys encoded on $N$ subcarriers within a single fiber channel [21, 22]. At present, two multi-carrier CV-QKD schemes have been proposed. One needs multiple CV-QKD transmitters, which is basically the same as deploying multiple CV-QKD systems [23]. The other is based on one transmitter with $N$ independent radio-frequency (RF) oscillators while it is difficult to realize due to the complicated electrical structure [24]. In fact, the multi-carrier CV-QKD scheme can be simply realized with one transmitter and one receiver by employing the orthogonal-frequency-division-multiplexing (OFDM) method [25-27], which is efficient and economical for practical deployment. Moreover, the SKR can be intuitively improved by increasing the carrier number $N$ for an OFDM-based multi-carrier CV-QKD. However, extra excess noise will be introduced in the imperfect multi-carrier quantum state preparation process, which will deteriorate the system performance for larger $N$. So, to effectively improve the system performance, there is a trade-off between the carrier number $N$ and the extra excess noise. The extra excess noise is mainly the modulation noise that originated from the IQ imbalance and intermodulation distortion. In [25], the IQ imbalance noise has been modeled and analyzed for the QPSK multi-carrier CV-QKD scheme. Nevertheless, there is still a lack of systematic modulation noise model for multi-carrier CV-QKD with arbitrary modulation protocol (e.g. QPSK, 256QAM or Gaussian protocol), which is seriously needed to evaluate the performance of the scheme.

In this paper, a systematic modulation noise model is proposed for the OFDM-based multi-carrier CV-QKD with arbitrary modulation protocol. Based on the proposed modulation noise model, the SKRs of the OFDM-based multi-carrier CV-QKD system with QPSK, 256QAM and Gaussian modulation protocols are quantitatively evaluated by combining the security analysis method of the single-carrier CV-QKD. Specifically, the modulation noise from IQ imbalance and third-order intermodulation distortion are quantitatively modeled and analyzed, which limits the performance of the system with $N$ subcarriers. Under practical system parameters, the simulation results show that the OFDM-based multi-carrier CV-QKD system can achieve much higher asymptotic SKR by increasing $N$ even with imperfect modulation compared with the single-carrier CV-QKD scheme. Moreover, a maximum total SKR for multi-carrier CV-QKD can be obtained by carefully optimizing $N$. Therefore, our work offers a promising way to further improve the SKR performance of quantum secure network.

## 2. Description of the multi-carrier CV-QKD scheme

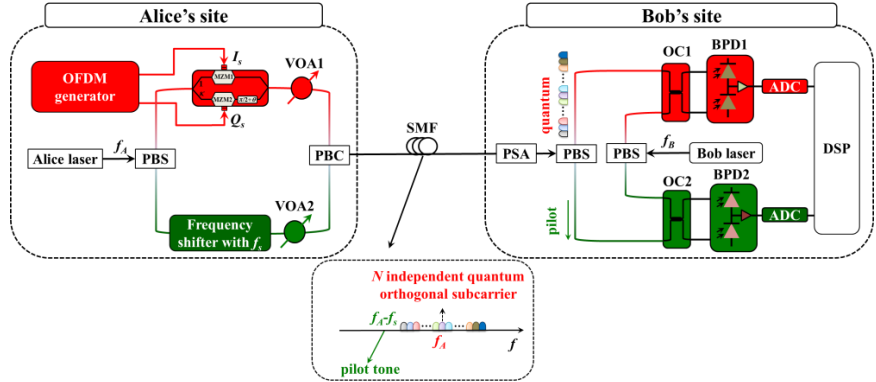

Fig. 1 Schematic diagram of the OFDM-based multi-carrier CV-QKD with local local oscillator. LD: laser diode; OFDM: orthogonal frequency division multiplexing; MZM: Mach-Zehnder modulator; PBS: polarization beam splitter; VOA, variable optical attenuator; PBC: polarization beam combiner; SMF: single mode fiber; PSA: polarization synthesis analyzer; OC: optical coupler; BPD: balanced photo-detector; DSP: digital signal processing

The working process of the OFDM-based multi-carrier CV-QKD scheme can be described as follows. (i) Quantum state preparation: Alice modulates $N$ independent random keys on a $N$ mode coherent state as $\bigotimes_{k=1}^{N}|X_k + P_k\rangle$ based on QPSK, 256QAM or Gaussian modulation CV-QKD protocol, and sends the prepared quantum state to Bob through the quantum channel. (ii) Quantum state measurement: Bob measures the received quantum state with coherent receiver (e.g. homodyne or heterodyne detection), and gets the measurement results $\vec{y} = \{y_k | k = 1,2, ... , N\}$ based on digital demodulation. Alice and Bob repeat steps (i) and (ii) for $n$ times. (iii) Post-processing: Alice and Bob perform sifting (if only Bob uses homodyne detection), parameter estimation, reverse reconciliation and privacy amplification independently for $N$ modes of encoded quantum states. The total asymptotic SKR between Alice and Bob with $N$ subcarriers can be written as:

$$R = \sum_{k=1}^{N} R_k, \quad R_k = \beta_k I(A_k:B_k) - \chi(B_k:E) \tag{1}$$

where $R_k$ is the asymptotic SKR of $k$-th mode, $I(A_k:B_k)$ is the Shannon mutual entropy between Alice and Bob for $k$-th mode, $\chi(B_k:E)$ is the quantum mutual entropy between Bob and Eve for $k$-th mode, and $\beta_k$ is the reconciliation efficiency for $k$-th mode.

In the following, we describe detailly the OFDM-based multi-carrier CV-QKD scheme with local local oscillator (LLO), as shown in Fig. 1. At Alice's site, a continuous optical carrier with central frequency $f_A$ that generated from Alice laser is split into two branches. In the upper branch, the optical carrier is modulated in an IQ modulator by OFDM signal $I_s$ and $Q_s$ generated from an OFDM generator. The processing routine of the OFDM generator is demonstrated in Fig. 2. Firstly, a serial of true random bit stream with high bit rate originated from quantum random number generator (QRNG) is converted to $N$ independent parallel bit streams by a serial-to-parallel conversion (S/P). Secondly, the $N$ parallel bit streams are mapped with arbitrary modulation distribution based on the chosen CV-QKD protocol, such as QPSK, 256QAM or Gaussian distribution. After inverse fast Fourier transform (IFFT), the processed parallel data is converted to the desired OFDM signal by a parallel-to-serial conversion (P/S), given by

$$I_s = \sum_{k=1}^{N} I_k \cos(2\pi f_k t) - \sum_{k=1}^{N} Q_k \sin(2\pi f_k t) \tag{2a}$$

$$Q_s = \sum_{k=1}^{N} I_k \sin(2\pi f_k t) + \sum_{k=1}^{N} Q_k \cos(2\pi f_k t) \tag{2b}$$

where $N$ is the total number of the generated subcarriers, $f_k = k\Delta f$ corresponds to the frequency of $k$-th subcarrier and $\Delta f$ denotes the frequency spacing between $N$ subcarriers. It is noteworthy from Fig. 2 that the cyclic prefix (CP) and training sequence (TS) are appended into the OFDM signal in processing routine for avoiding mutual crosstalk between subcarriers in fiber channel and achieving the data-aided time domain equalization at Bob's site.

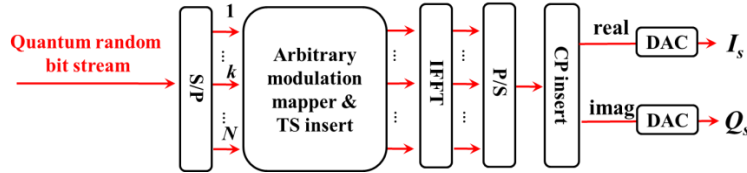

Fig. 2 The data processing of the OFDM generator. S/P: serial-to-parallel conversion; TS: training sequence; IFFT: inverse fast Fourier transform; P/S: parallel-to-serial conversion; CP: cyclic prefix; DAC: digital-to-analog converter.

In the upper branch at Alice's site in Fig. 1, the OFDM-based optical signal after IQ modulation is attenuated by a variable optical attenuator (VOA) to generate the encoded multi-carrier quantum state

$$\left| X_{sig} + jP_{sig} \right\rangle = \otimes_{k=1}^{N} \left| X_k + jP_k + \Delta X_k + j\Delta P_k \right\rangle \tag{3}$$

where $X_k$ and $P_k$ are the expected modulated quadrature components for the $k$-th subcarrier. $\Delta X_k$ and $\Delta P_k$ are the extra modulation noise on $X_k$ and $P_k$, respectively, which mainly come from IQ imbalance and intermodulation distortion in the multi-carrier IQ modulation process. The detailed theoretical model of the modulation process is shown in next section. In the lower branch, the optical carrier is sent into a frequency shifter (e.g. MZM with carrier suppression double sideband modulation) with $f_s$ for generating the desired pilot tone with frequency $f_A$-$f_s$ [28]. The generated pilot tone is used for eliminating the fast-drift phase noise in the LLO-CV-QKD system. Finally, the multi-carrier quantum signal and the pilot tone are co-transmitted into a single mode fiber (SMF) channel with different frequency bands and orthogonal polarization states for avoiding the crosstalk between them.

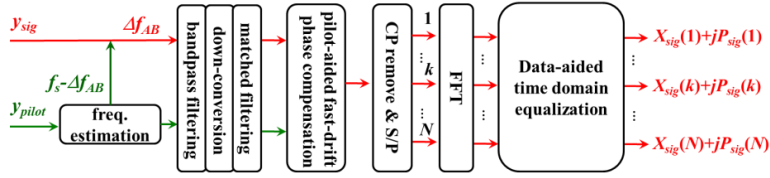

Fig. 3 The DSP routine of the OFDM-based multi-carrier LLO-CV-QKD scheme. CP: cyclic prefix; S/P: serial-to-parallel conversion; FFT: fast Fourier transform.

At Bob's site, the received quantum signal and pilot tone are separated by a polarization beam splitter (PBS) after the polarization deviation correction with a polarization synthesis analyzer (PSA). The separated quantum signal and pilot tone are independently coupled into two balanced photo-detectors (BPDs) with their corresponding LLO signals of frequency $f_B$ supplied by the Bob laser. The detected quantum signal and pilot tone are analog-to-digital converted for the subsequent DSP, respectively. As is shown in Fig. 3, the frequency $f_s$-$\Delta f_{AB}$ is firstly efficiently estimated from the pilot tone, with which the central frequency $\Delta f_{AB}$ of the OFDM-based quantum signal can be obtained. Secondly, the OFDM-based quantum signal $y_{sig}$ and the pilot tone $y_{pilot}$ are bandpass filtered, down-converted and matched filtered for recovering their quadrature components. Thirdly, the fast-drift phase noise of the OFDM-based quantum signal can be compensated based on the pilot-assisted channel equalization method, which is recently applied in optical fiber LLO-CV-QKD system [29]. After removing CP, the serial OFDM-based quantum signal is converted to $N$ parallel signals and then transformed to $N$ independent subcarriers by FFT. Subsequently, the phase noises of $N$ independent subcarriers are further compensated based on data-aided time domain equalization method, respectively. Finally, a high-efficient post-processing setup for the multi-carrier CV-QKD is required to extract $N$ independent secure keys.

## 3. Modulation noise model of the multi-carrier CV-QKD

In the OFDM-based multi-carrier modulation process, the preparation accuracy of the multi-carrier quantum state will suffer from the IQ imbalance and intermodulation distortion. Therefore, a detailed modulation noise model of the multi-carrier CV-QKD scheme should be derived to effectively evaluate the SKR.

The OFDM-based multi-carrier quantum signal after the IQ modulation can be written as:

$$E_{sig}(t) = 2A_{sig}e^{j2\pi f_A t}\left\{G_1\sin\left[\sum_{k=1}^{N}\gamma_k\cos(2\pi f_k t+\varphi_k)\right] + jG_2\sin\left[\sum_{k=1}^{N}\gamma_k\sin(2\pi f_k t+\varphi_k)\right]\right\} \quad (4)$$

with

$$\gamma_k = \mu_k\sqrt{I_k^2+Q_k^2} \quad (5a)$$

$$\cos\varphi_k = \frac{I_k}{\sqrt{I_k^2+Q_k^2}} \quad (5b)$$

where the IQ imbalance factor $G_1=[1+\kappa_k\exp(j\theta_k)]/2$ and $G_2=[1+\kappa_k\exp(-j\theta_k)]/2$, and $\kappa_k$ and $\theta_k$ are the gain imbalance and quadrature skew of the $k$-th subcarrier modulation, respectively. $\mu_k=V_k\pi/V_\pi/2$ is the modulation index of the $k$-th subcarrier, and $V_k$ and $V_\pi$ are the driving amplitude and half-wave voltage of the employed I/Q modulator, respectively. $A_{sig}$ is the amplitude of the quantum signal. Significantly, under IQ balance ($\kappa_k=1$ and $\theta_k=0$) and no intermodulation distortion, Eq. (4) can be simplified as:

$$E_{sig}(t) = 2A_{sig}e^{j2\pi f_A t}\sum_{k=1}^{N}\gamma_k e^{j(2\pi f_k t+\varphi_k)} = 2A_{sig}e^{j2\pi f_A t+j2\pi f_k t}\otimes_{k=1}^{N}|I_k+jQ_k\rangle \quad (6)$$

Comparing Eq. (3) and Eq. (6), one can see that the expected modulated quadrature $X_k$ and $P_k$ of the $k$-th subcarrier are determined by the random variables $I_k$ and $Q_k$ with arbitrary modulation distribution.

In order to derive modulation noise model for $\Delta X_k$ and $\Delta P_k$ shown in Eq. (3), we apply the Jacobi-Anger expansion [30] into Eq. (4), which can be expanded with the $p$th-order Bessel function of the first kind $J_p(\cdot)$ ($p=0, \pm 1, \pm 2, \pm 3, \cdots$) as follows:

$$E_{sig}(t) = 2A_{sig}e^{j2\pi f_A t}\left\{\begin{array}{l} G_1 \sum_{p_1=-\infty}^{+\infty} J_{p_1}(\gamma_1) \cdots \sum_{p_k=-\infty}^{+\infty} J_{p_k}(\gamma_k) \cdots \sum_{p_N=-\infty}^{+\infty} (\gamma_N) J_{p_N} \cdot \\ \sin\left[\begin{array}{l}\left(2\pi f_1 t p_1 + p_1\varphi_1 + p_1\frac{\pi}{2}\right) + \cdots + \left(2\pi f_k t p_k + p_k\varphi_k + p_k\frac{\pi}{2}\right) \\ + \cdots + \left(2\pi f_N t p_N + p_N\varphi_N + p_N\frac{\pi}{2}\right)\end{array}\right] \\ + jG_2 \sum_{p_1=-\infty}^{+\infty} J_{p_1}(\gamma_1) \sum_{p_k=-\infty}^{+\infty} J_{p_k}(\gamma_2) \cdots \sum_{p_N=-\infty}^{+\infty} J_{p_N}(\gamma_N) \cdot \\ \sin\left[\begin{array}{l}(2\pi f_1 t p_1 + p_1\varphi_1) + \cdots + (2\pi f_k t p_k + p_k\varphi_k) \\ + \cdots + (2\pi f_N t p_N + p_N\varphi_N)\end{array}\right]\end{array}\right\} \quad (7)$$

From Eq. (7), we can quantify the frequency component at $f_0+f_k$ for expressing the $k$-th subcarrier in consideration of the IQ imbalance and the third-order intermodulation distortion, which is given by

$$E_{sig(k)}(t) = 2A_{sig}e^{j\pi f_A t} \cdot \left\{\begin{array}{l} J_1(\gamma_k)\left[(G_1+G_2)e^{j(2\pi f_k t+\varphi_k)} + (G_1-G_2)e^{-j(2\pi f_k t+\varphi_k)}\right] \\ -\sum_{\substack{m,n=1 \\ m\neq n\neq k \\ 2m+n=k}}^{N} J_2(\gamma_m)J_1(\gamma_n)\left[(G_1+G_2)e^{-j(2\pi f_k t+2\varphi_m+\varphi_n)} + (G_1-G_2)e^{j(2\pi f_k t+2\varphi_m+\varphi_n)}\right] \\ -\sum_{\substack{m,s=1 \\ m\neq n\neq k \\ 2m-n=k}}^{N} J_2(\gamma_m)J_1(\gamma_n)\left[(G_1+G_2)e^{j(2\pi f_k t+2\varphi_m-\varphi_n)} + (G_1-G_2)e^{-j(2\pi f_k t+2\varphi_m-\varphi_n)}\right] \\ -\sum_{\substack{m,s,l=1 \\ m\neq n\neq l\neq k \\ m+n+l=k}}^{N} J_1(\gamma_m)J_1(\gamma_n)J_1(\gamma_l)\left[(G_1+G_2)e^{-j(2\pi f_k t+\varphi_m+\varphi_n+\varphi_l)} + (G_1-G_2)e^{j(2\pi f_k t+\varphi_m+\varphi_n+\varphi_l)}\right] \\ -\sum_{\substack{m,n,l=1 \\ m\neq n\neq l\neq k \\ m+n-l=k}}^{N} J_1(\gamma_m)J_1(\gamma_n)J_1(\gamma_l)\left[(G_1+G_2)e^{j(2\pi f_k t+\varphi_m+\varphi_n-\varphi_l)} + (G_1-G_2)e^{-j(2\pi f_k t+\varphi_m+\varphi_n-\varphi_l)}\right] \\ -\sum_{\substack{m,n,l=1 \\ m\neq n\neq l\neq k \\ m-n-l=k}}^{N} J_1(\gamma_m)J_1(\gamma_n)J_1(\gamma_l)\left[(G_1+G_2)e^{-j(2\pi f_k t+\varphi_m-\varphi_n-\varphi_l)} + (G_1-G_2)e^{j(2\pi f_k t+\varphi_m-\varphi_n-\varphi_l)}\right] \end{array}\right\} \quad (8)$$

Under small signal condition of $J_1(x) \approx x/2$ and $J_2(x) \approx x^2/4$, the quadrature $X_{sig(k)}$ of the $k$-th subcarrier can be obtained from Eq. (8) and expressed as:

$$X_{sig(k)} = X_k + \Delta X_k = 2A_{sig}\,\text{real}\left(\gamma_k e^{j\varphi_k}\right)$$

$$+2A_{sig}\,\text{real}\left\{\begin{array}{l} \frac{\gamma_k}{2}\left[(G_1+G_2-2)e^{j\varphi_k} + (G_1-G_2)e^{-j\varphi_k}\right] \\ -\frac{M_1(N,k)\gamma_m^2\gamma_n}{8}\left[(G_1+G_2)e^{-j(2\varphi_m+\varphi_n)} + (G_1-G_2)e^{j(2\varphi_m+\varphi_n)}\right] \\ -\frac{M_2(N,k)\gamma_m^2\gamma_n}{8}\left[(G_1+G_2)e^{j(2\varphi_m-\varphi_n)} + (G_1-G_2)e^{-j(2\varphi_m-\varphi_n)}\right] \\ -\frac{W_1(N,k)\gamma_m\gamma_n\gamma_l}{8}\left[(G_1+G_2)e^{-j(\varphi_m+\varphi_n+\varphi_l)} + (G_1-G_2)e^{j(\varphi_m+\varphi_n+\varphi_l)}\right] \\ -\frac{W_2(N,k)\gamma_m\gamma_n\gamma_l}{8}\left[(G_1+G_2)e^{j(\varphi_m+\varphi_n-\varphi_l)} + (G_1-G_2)e^{-j(\varphi_m+\varphi_n-\varphi_l)}\right] \\ -\frac{W_3(N,k)\gamma_m\gamma_n\gamma_l}{8}\left[(G_1+G_2)e^{-j(\varphi_m-\varphi_n-\varphi_l)} + (G_1-G_2)e^{j(\varphi_m-\varphi_n-\varphi_l)}\right] \end{array}\right\} \quad (9)$$

where $M_1(N, k)$ and $M_2(N, k)$ are the combinatorial numbers of the condition $2m+n=k$ and $2m-$

$n=k$, respectively. $W_1(N, k)$, $W_2(N, k)$ and $W_3(N, k)$ are the combinatorial numbers of the condition $m+n+l=k$, $m+n-l=k$ and $m-n-l=k$, respectively. In Eq. (9), $X_k$ is the $k$-th subcarrier component under the condition of IQ balance ($\kappa_k=1$ and $\theta_k=0$) and no intermodulation distortion [$M_1(N, k)=M_2(N, k)=W_1(N, k)=W_2(N, k)=W_3(N, k)=0$], which can be expressed as

$$X_k = 2A_{sig}\mu_k I_k \qquad (10)$$

So, the expected modulation variance of the $k$-th subcarrier can be expressed as $V_A = \langle X_k^2 \rangle = 4A_{sig}^2 \mu_k^2 \langle I_k^2 \rangle$. In Eq. (9), $\Delta X_k$ is the extra modulation noise introduced from the IQ imbalance and third-order intermodulation distortion, which includes three parts

$$\Delta X_k = \Delta X_{k1} + \Delta X_{k2} + \Delta X_{k3} \qquad (11)$$

with

$$\Delta X_{k1} = A_{sig}\mu_k \left[ (\kappa_k \cos\theta_k - 1)I_k + (\kappa_k \sin\theta_k) Q_k \right] \qquad (12a)$$

$$\Delta X_{k2} = \frac{1}{4} A_{sig} \mu_m^2 \mu_n \left\{ (1+\kappa_k \cos\theta_k) \begin{bmatrix} [M_1(N,k)+M_2(N,k)]Q_m^2 I_n \\ +2[M_1(N,k)-M_2(N,k)]I_m Q_m Q_n \\ -[M_1(N,k)+M_2(N,k)]I_m^2 I_n \end{bmatrix} + (\kappa_k \sin\theta_k) \begin{bmatrix} [M_1(N,k)+M_2(N,k)]I_m^2 Q_n \\ +2[M_1(N,k)-M_2(N,k)]I_m Q_m I_n \\ -[M_1(N,k)+M_2(N,k)]Q_m^2 Q_n \end{bmatrix} \right\} \qquad (12b)$$

$$\Delta X_{k3} = \frac{1}{4} A_{sig} \mu_m \mu_n \mu_l \left\{ (1+\kappa_k \cos\theta_k) \begin{bmatrix} [W_1(N,k)-W_2(N,k)+W_3(N,k)]I_m Q_n Q_l \\ +[W_1(N,k)-W_2(N,k)-W_3(N,k)]Q_m I_n Q_l \\ +[W_1(N,k)+W_2(N,k)-W_3(N,k)]Q_m Q_n I_l \\ -[W_1(N,k)+W_2(N,k)+W_3(N,k)]I_m I_n I_l \end{bmatrix} + (\kappa_k \sin\theta_k) \begin{bmatrix} [W_1(N,k)+W_2(N,k)-W_3(N,k)]I_m I_n Q_l \\ +[W_1(N,k)-W_2(N,k)-W_3(N,k)]I_m Q_n I_l \\ +[W_1(N,k)-W_2(N,k)+W_3(N,k)]Q_m I_n I_l \\ -[W_1(N,k)+W_2(N,k)+W_3(N,k)]Q_m Q_n Q_l \end{bmatrix} \right\} \qquad (12c)$$

where $\Delta X_{k1}$ is modulation noise resulted from the IQ imbalance. The modulation noise $\Delta X_{k2}$ and $\Delta X_{k3}$ are originated from the third-order intermodulation with the frequency relationships ($2f_m\text{-}f_n=f_k$ and $2f_m+f_n=f_k$) and ($f_m+f_n+f_l=f_k$, $f_m+f_n\text{-}f_l=f_k$ and $f_m\text{-}f_n\text{-}f_l=f_k$), respectively.

It is obvious that the quadrature $I_g$ and $Q_g$ ($g=k, m, n, l$) of the $g$-th subcarrier are mutually random and independent. One can define the variance $\langle I_g^2 \rangle = \langle Q_g^2 \rangle = \sigma_1^2$ and $\langle I_g^4 \rangle = \langle Q_g^4 \rangle = \sigma_2^2$. According to Eq. (11), the variances of $\Delta X_k$ can be expressed as

$$\langle \Delta X_k^2 \rangle = \langle \Delta X_{k1}^2 \rangle + \langle \Delta X_{k2}^2 \rangle + \langle \Delta X_{k3}^2 \rangle \qquad (13)$$

with

$$\langle \Delta X_{k1}^2 \rangle = A_{sig}^2 \mu_k^2 \sigma_1^2 \left( \kappa_k^2 + 1 - 2\kappa_k \cos\theta_k \right) \qquad (14a)$$

$$\langle \Delta X_{k2}^2 \rangle = \frac{A_{sig}^2 \mu_m^4 \mu_n^2}{8}\left(1+2\kappa_k \cos\theta_k + \kappa_k^2\right) \left\{ \begin{array}{l} [M_1(N,k)+M_2(N,k)]^2 \sigma_2^2 \sigma_1^2 \\ +2[M_1(N,k)-M_2(N,k)]^2 \sigma_1^6 \end{array} \right\} \qquad (14b)$$

$$\langle \Delta X_{k3}^2 \rangle = \frac{A_{sig}^2 \mu_m^2 \mu_n^2 \mu_l^2}{4}\left(1+2\kappa_k \cos\theta_k + \kappa_k^2\right)\left[W_1^2(N,k)+W_2^2(N,k)+W_3^2(N,k)\right]\sigma_1^6 \qquad (14c)$$

At the same time, $\mu_k$, $\mu_m$, $\mu_n$ and $\mu_l$ are the modulation indexes of the $k$, $m$, $n$, $l$-th subcarrier modulation, respectively, which are approximately equal to each other in the multi-carrier modulation process. Based on the above equations, we can use $\mu_k$ to replace $\mu_m$, $\mu_n$ and $\mu_l$, and

simplify the modulation noise of the $k$-th subcarrier with the actual modulation variance $V_A$, given by

$$\varepsilon_{mod}(k) = \langle \Delta X_k^2 \rangle = \frac{V_A}{4}\left(\kappa_k^2 + 1 - 2\kappa_k \cos\theta_k\right) \tag{15}$$

$$+ \frac{V_A \mu_k^4}{32}\left(1 + 2\kappa_k \cos\theta_k + \kappa_k^2\right)\left\{\begin{array}{l}[M_1(N,k)+M_2(N,k)]^2 \sigma_2^2 \\ +2\left\{\begin{array}{l}[M_1(N,k)-M_2(N,k)]^2 \\ +W_1^2(N,k)+W_2^2(N,k)+W_3^2(N,k)\end{array}\right\}\sigma_1^4\end{array}\right\}$$

One can see from Eq. (15) that the modulation noise of the $k$-th subcarrier in the multi-carrier CV-QKD system is related with the modulation index $\mu_k$, the IQ imbalance factors ($\kappa_k$ and $\theta_k$) and the third-order intermodulation factors [$M_1(N, k)$, $M_2(N, k)$, $W_1(N, k)$, $W_2(N, k)$ and $W_3(N, k)$]. Moreover, the modulation noise is also determined by the variance $\sigma_1^2$ and $\sigma_2^2$ related with different modulation distributions (e.g. QPSK, 256QAM or Gaussian modulation) in the quantum state preparation process.

## 4. Performance evaluation of multi-carrier CV-QKD

It is well known that the modulation noise is estimated as a part of excess noise at Alice's site [31]. In this case, the extra modulation noise $\varepsilon_{mod}(k)$ of the $k$-th subcarrier in the multi-carrier CV-QKD will cause an increment of the excess noise compared to a single-carrier CV-QKD, and the total excess noise $\varepsilon_{multi}(k)$ of the $k$-th subcarrier can be expressed as

$$\varepsilon_{multi}(k) = \varepsilon_{mod}(k) + \varepsilon_{single} \tag{16}$$

where $\varepsilon_{single}$ is the excess noise of a single-carrier CV-QKD scheme without the modulation noise. Based on the security analysis method of the single-carrier CV-QKD scheme, the SKR performances of the multi-carrier CV-QKD with QPSK, 256 QAM and Gaussian modulation can be evaluated with modulation noise $\varepsilon_{mod}(k)$ simulated from the proposed modulation noise model.

### 4.1 QPSK modulation protocol

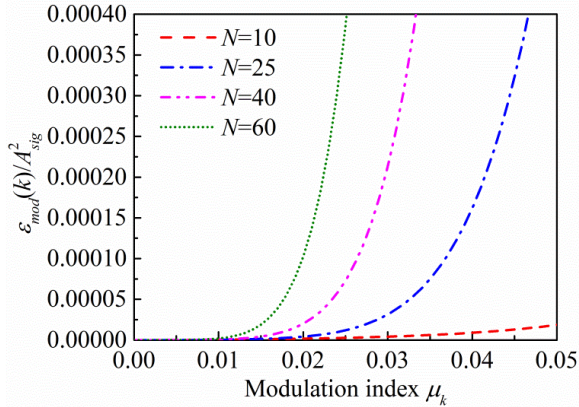

Fig. 4 Simulated $\varepsilon_{mod}(k)/A^2_{sig}$ of the $k$-th subcarrier at different modulation index $\mu_k$ with $N$=10, 25, 40 and 60, respectively.

For the QPSK modulation multi-carrier CV-QKD scheme, the variance $\sigma_1^2$ and $\sigma_2^2$ are determined to be 1 because the random variables $I_g$ and $Q_g$ ($g=k, m, n, l$) follow mutually independent binomial distribution (-1, 1). Under the total carrier number $N$=10, 25, 40 and 60, the gain imbalance $\kappa_k$=0.98 and the quadrature skew $\theta_k=\pi/50$, the modulation noise ratio $\varepsilon_{mod}(k)/A^2_{sig}$ of the $k$-th subcarrier at different modulation index $\mu_k$ are calculated according to Eq. (15) and shown in Fig. 4. It is seen from Fig. 4 that $\varepsilon_{mod}(k)/A_{sig}^2$ of the $k$-th subcarrier is an increasing function of modulation index $\mu_k$. Therefore, it is better to choose a lower modulation index $\mu_k$ to reduce the modulation noise especially for the larger total carrier number $N$ in practical experiment, e.g. the modulation noise ratio for $N$=60 is close to 0 when $\mu_k$=0.01.

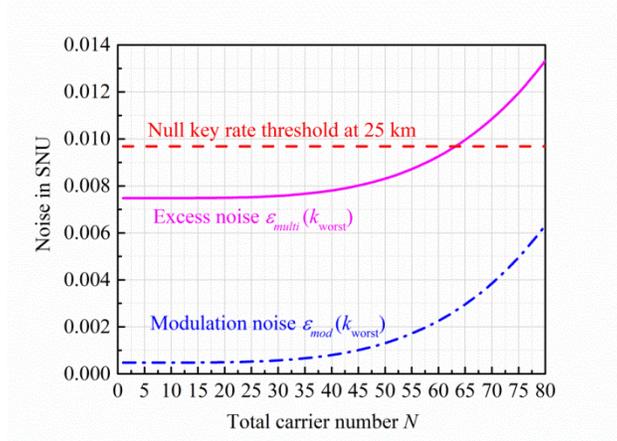

Fig. 5 Calculated noise of the $k_{worst}$-th subcarrier at different total carrier number $N$ in the QPSK multi-carrier CV-QKD. The blue dash dot line denotes the worst modulation noise $\varepsilon_{mod}(k_{worst})$, the magenta solid line corresponds to the excess noise $\varepsilon_{multi}(k_{worst})$ and the red dash line represents the null key rate threshold at 25 km.

With system parameters $\mu_k$=0.01, $\kappa_k$=0.98 and $\theta_k$=$\pi$/50, the worst modulation noise $\varepsilon_{mod}(k_{worst})$ of the $k_{worst}$-th subcarrier in different $N$ carriers is numerically obtained as shown in Fig. 5. According to our previous QPSK single-carrier CV-QKD experimental results [10], the single-carrier excess noise $\varepsilon_{single}$ can be controlled to be about 0.007 shot noise unit (SNU) with optimized modulation variance $V_A$=0.45 SNU. Suppose the $k$-th subcarrier share the same excess noise performance as the single-carrier CV-QKD scheme, except with the extra modulation noise induced in the multi-carrier quantum state prepration process. Then, the excess noise $\varepsilon_{multi}(k_{worst})$ of the $k_{worst}$-th subcarrier in the QPSK multi-carrier CV-QKD scheme can be calculated based on Eq. (16), which is shown in Fig. 5. The excess noise $\varepsilon_{multi}(k_{worst})$ of the $k_{worst}$-th subcarrier gradually increases with the total carrier number $N$, which should be set to below 60 for avoiding null key rate at 25 km transmission distance.

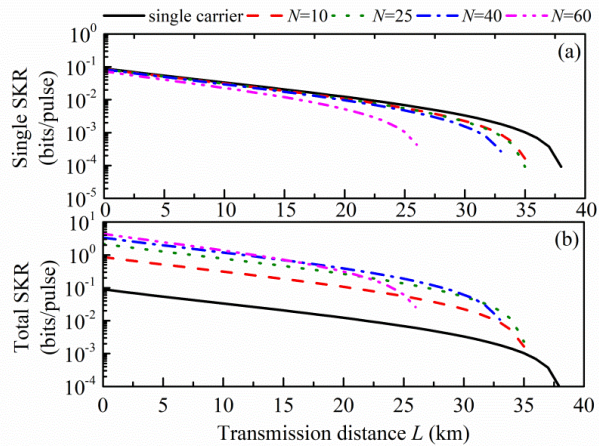

Fig. 6 The SKR of the QPSK multi-carrier CV-QKD with $N$=10, 25, 40 and 60, respectively. (a) Single SKR of the $k_{worst}$-th subcarrier, (b) total SKR of the QPSK multi-carrier CV-QKD.

Under the modulation variance $V_A$=0.45 SNU, the reconciliation efficiency $\beta_k$=95% ($k$=1, 2, …, $N$), and the excess noise $\varepsilon_{multi}(k_{worst})$ calculated by Eq. (16), the single SKR $R_{k_{worst}}$ for the $k_{worst}$-th subcarrier is evaluated by using the semidefinite programming (SDP) security analysis method [32, 33], as shown in Fig. 6(a). In fact, the SKR can be further improved by employing the security analysis method in [34] for QPSK CV-QKD protocol. Note that the modulation noise $\varepsilon_{mod}(k_{worst})$ of the $k_{worst}$-th subcarrier is the worst in $N$ subcarriers. So, the total SKR for the multi-carrier CV-QKD with $N$ carriers can be obtained as $R \geq NR_{k_{worst}}$ according to Eq. (1), as shown in Fig. 6(b). Moreover, the multi-carrier SKR gain that is defined as the ratio between the total SKR of multi-carrier CV-QKD with extra modulation noise and that of single-carrier CV-QKD without modulation noise is shown in Fig.7, which indicates that the SKR of the multi-carrier CV-QKD can be greatly improved by increasing $N$. It is noteworthy that the multi-carrier SKR gain drastically drops with the increase of total carrier number $N$ in long distance transmission due to the extra modulation noise. Besides, the optimal carrier number $N$ can be numerically calculated based on the proposed modulation noise model as shown in Fig. 8, where the optimal $N$=61, 56 and 43 for $L$=5 km, 10 km and 25 km, respectively.

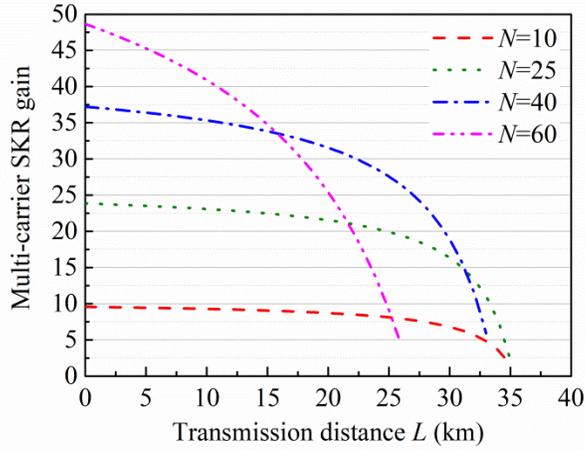

Fig. 7 Calculated multi-carrier SKR gains for the QPSK multi-carrier CV-QKD with $N$=10, 25, 40 and 60, respectively.

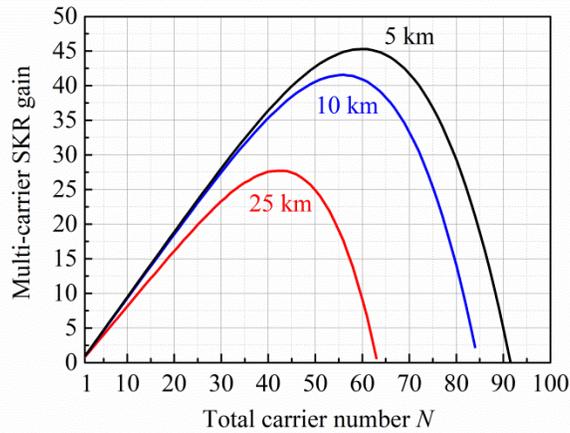

Fig. 8 Calculated QPSK multi-carrier SKR gains at different total carrier number $N$ with transmission distance $L$=5 km, 10 km, 25 km, respectively.

## 4.2 256QAM modulation protocol

The SKR performance of multi-carrier CV-QKD with 256QAM discrete modulation protocol is evaluated in a similar way based on the proposed modulation noise model with system parameters $\mu_k=0.01$, $\kappa_k=0.98$ and $\theta_k=\pi/50$. For 256QAM modulation, the random variables $I_g$ and $Q_g$ ($g=k, m, n, l$) obey mutually independent discrete Maxwell-Boltzmann distribution, and the corresponding probability density function is given by [11]

$$\rho_{x_i} = \frac{\exp(-\nu x_i^2)}{\sum_{i=1}^{16}\exp(-\nu x_i^2)} \tag{17}$$

where $x_i$ is normalized within (-1 1). $\nu$ is a positive free parameter whose optimal value is about 0.04. Therefore, the variance $\sigma_1^2$ of the quadrature $I_g$ and the variance $\sigma_2^2$ of the quadrature $I_g^2$ can be obtained as, respectively

$$\sigma_1^2 = \sum_{i=1}^{16}\left[x_i - E(x_i)\right]^2 \rho_{x_i} \tag{18a}$$

$$\sigma_2^2 = \sum_{i=1}^{16}\left[x_i^2 - E(x_i^2)\right]^2 \rho_{x_i}^2 \tag{18b}$$

with $\sigma_1^2=0.37$ and $\sigma_2^2=0.11$ applied in Eq. (15). So, the single SKR of the $k_\text{worst}$-th subcarrier in 256QAM multi-carrier CV-QKD can be evaluated by using the improved SDP method in [33], as is shown in Fig. 9(a). Moreover, the modulation variance $V_A=5$ SNU, the reconciliation efficiency $\beta_k=95\%$, the quantum efficiency $\eta=0.56$, the electronic noise $\upsilon_{ele}=0.15$ SNU and the excess noise $\varepsilon_{single}=0.03$ SNU are chosen in the following simulations referring to our previous 256QAM single-carrier CV-QKD experiment [11]. Note that the single-carrier excess noise $\varepsilon_{single}=0.03$ SNU has excluded the IQ imbalance noise by DSP in the single-carrier CV-QKD experiment [11]. Furthermore, the corresponding total SKR and the multi-carrier SKR gain are respectively given in Fig. 9(b) and Fig. 10 for verifying the performance of 256QAM multi-carrier CV-QKD. Besides, the multi-carrier SKR gains with different $N$ are calculated in Fig. 11, where the optimal $N=$ 96, 85 and 72 for $L=25$ km, 50 km and 100 km, respectively.

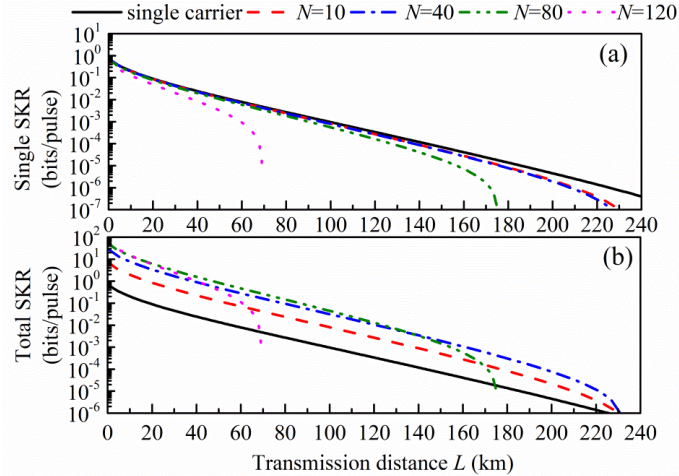

Fig. 9 The SKR of the 256QAM multi-carrier CV-QKD with $N=10$, 40, 80 and 120, respectively. (a) Single SKR of the $k_\text{worst}$-th subcarrier, (b) total SKR of the 256QAM multi-carrier CV-QKD.

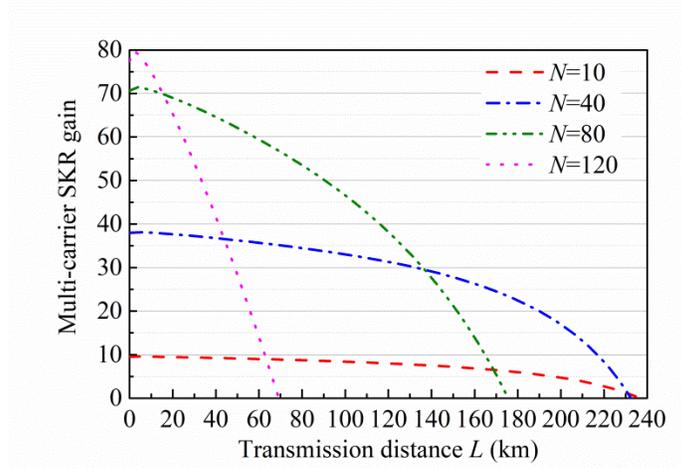

Fig. 10 Calculated multi-carrier SKR gains for the 256QAM multi-carrier CV-QKD with $N$=10, 40, 80 and 120, respectively.

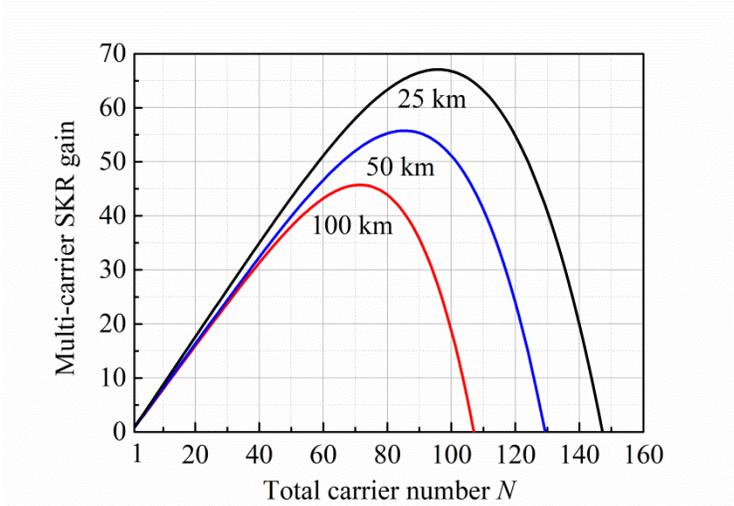

Fig. 11 Calculated 256QAM multi-carrier SKR gains at different total carrier number $N$ with transmission distance $L$=25 km, 50 km and 100 km, respectively.

### 4.3 Gaussian modulation protocol

We also estimate the SKR performance of the multi-carrier CV-QKD with Gaussian modulation protocol based on the proposed modulation noise model with system parameters $\mu_k$=0.01, $\kappa_k$=0.98 and $\theta_k=\pi/50$. The random variables $I_g$ and $Q_g$ ($g=k, m, n, l$) are mutually independent and follow the normal Gaussian distribution $N(0, \sigma_1^2)$ with amplitude normalization (-1, 1), whose with probability density function is given by

$$f(x) = \frac{1}{\sqrt{2\pi}\sigma_1} e^{-\frac{x^2}{2\sigma_1^2}} \tag{19}$$

and the corresponding random variables $I_g^2$ and $Q_g^2$ follow the Gaussian distribution $N(\sigma_1^2, 2\sigma_1^4)$, which can be verified as

$$E(x^2) = \int_{-\infty}^{+\infty} x^2 f(x) dx = \sigma_1^2 \tag{20a}$$

$$\sigma_2^2 = E(x^4) - \left[E(x^2)\right]^2 = \int_{-\infty}^{+\infty} x^4 f(x) dx - \sigma_1^4 = 2\sigma_1^4 \tag{20b}$$

Therefore, the variance $\sigma_1^2=1/9$ and $\sigma_2^2=2/81$ can be obtained when the standard normal distribution $N(0, \sigma^2=1)$ is normalized to (-1, 1) from the amplitude section (-3σ, 3σ). Similarly, applying $\sigma_1^2$ and $\sigma_2^2$ in Eq. (15), the single SKR of the $k_{worst}$-th subcarrier for the Gaussian modulated multi-carrier CV-QKD can be evaluated based on the no-switch protocol security analysis method [35], as shown in Fig. 12(a). In the SKR calculation, the modulation variance $V_A$=5 SNU, the quantum efficiency $\eta$=0.56, the electronic noise $\upsilon_{ele}$=0.1 SNU, the reconciliation efficiency $\beta_k$=95% and the excess noise $\varepsilon_{single}$=0.03 SNU are chosen as in [28]. Similarly, the corresponding total SKR and multi-carrier SKR gain are given in Fig. 12(b) and Fig. 13, respectively, which also verifies that the SKR can be greatly increased based on the multi-carrier CV-QKD scheme. Moreover, the multi-carrier SKR gains at different $N$ are demonstrated Fig. 14, where the optimal $N$=155, 130 and 110 for $L$=50 km, 100 km and 150 km, respectively.

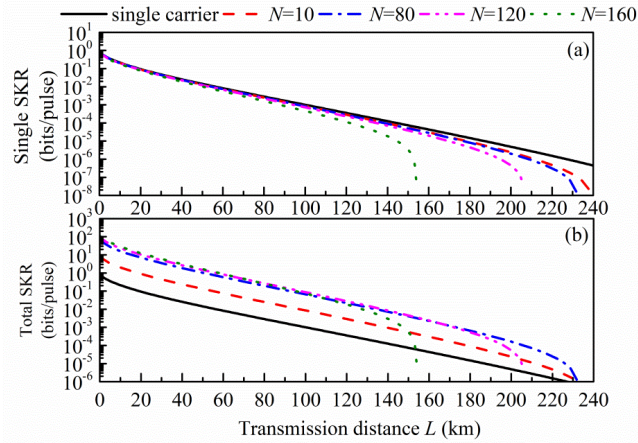

Fig. 12 The SKR of the Gaussian modulated multi-carrier CV-QKD with $N$=10, 80, 120 and 160, respectively. (a) Single SKR of the $k_{worst}$-th subcarrier, (b) total SKR of the Gaussian modulated multi-carrier CV-QKD.

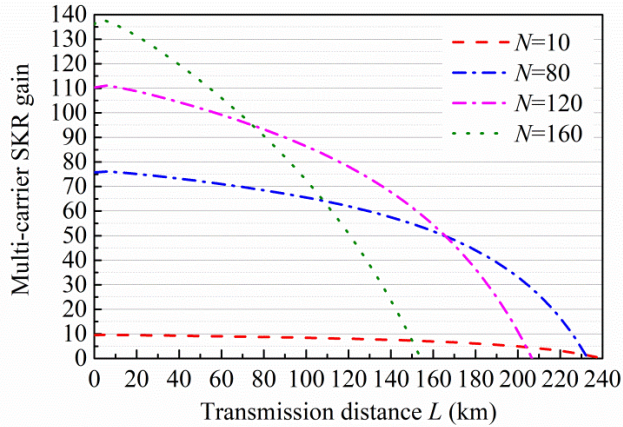

Fig. 13 Calculated multi-carrier SKR gains for Gaussiam modulated multi-carrier CV-QKD with $N$=10, 80, 120 and 160, respectively

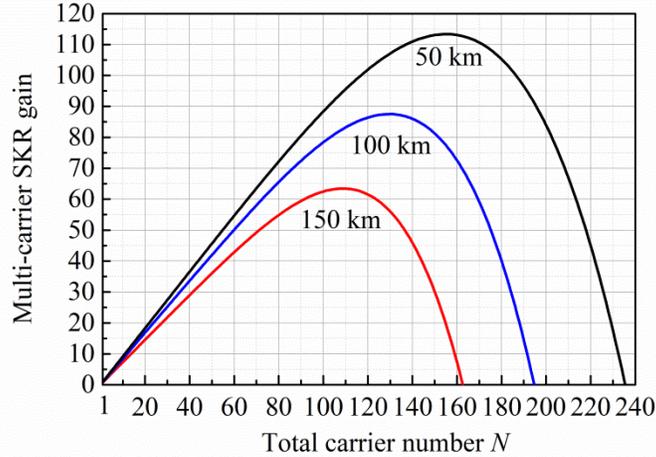

Fig. 14 Calculated Gaussian modulated multi-carrier SKR gains at different total carrier number *N* with transmission distance *L*=50 km, 100 km and 150 km, respectively.

## 5. Conclusion

We have demonstrated a systematic modulation noise model for OFDM-based multi-carrier CV-QKD with arbitrary modulation protocol. Specifically, the extra modulation noise including the IQ imbalance noise and the third-order intermodulation noise in the multi-carrier quantum state preparation is theoretically modeled and analyzed, which limits the SKR performance of system with total carrier number *N*. Under considering the extra modulation noise, the SKR performance of the OFDM-based multi-carrier LLO-CV-QKD system with QPSK, 256QAM and Gaussian modulation protocol can be quantitatively evaluated by combining the security analysis method of the single-carrier CV-QKD system. Under practical system parameters, our simulation results show that the OFDM-based multi-carrier CV-QKD scheme can greatly improve the asymptotic SKR by increasing *N* even with imperfect modulations. Moreover, a maximum total SKR can be obtained by carefully optimizing *N*. Compared with the single-carrier CV-QKD scheme, our work offers a promising way to significantly improve the performance of CV-QKD. In the future, other noises from the channel and receiver in practical experiment should be further considered to evaluate the performance of multi-carrier CV-QKD system, such as the fiber channel crosstalk noise between these subcarriers.

## Funding

This work was supported in part by the National Key Research and Development Program of China (Grant No. 2020YFA0309704), the National Natural Science Foundation of China (Grant Nos U19A2076, 62101516, 62171418, 62201530 and 61901425), the Technology Innovation and Development Foundation of China Cyber Security (Grant No. JSCX2021JC001), the Chengdu Major Science and Technology Innovation Program (Grant No. 2021-YF08-00040-GX), the Chengdu Key Research and Development Support Program (Grants Nos 2021-YF05-02430-GX and 2021-YF09-00116-GX), the Sichuan Science and Technology Program (Grants Nos 2022YFG0330 and 2021YJ0313), the Foundation of Science and Technology on Communication Security Laboratory (Grant No. 61421030402012111), and the Major Project of the Department of Science and Technology of Sichuan (2022ZDZX0009).

## Disclosures

The authors declare that there are no conflicts of interest related to the article.